\documentstyle[11pt,newpasp,twoside]{article}

\newcommand{\beq}{\begin{equation}}
\newcommand{\eeq}{\end{equation}}
\def\beqa{\begin{eqnarray}}
\def\eeqa{\end{eqnarray}}

\input psfig

\def\edcomment#1{\iffalse\marginpar{\raggedright\sl#1\/}\else\relax\fi}
\reversemarginpar

\begin{document}

\title{The Cosmological Tests}
\author{P. J. E. Peebles}
\affil{Joseph Henry Laboratories, Princeton University,
Princeton, NJ 08544}

\begin{abstract}
Recent observational advances have considerably improved the
cosmological tests, adding to the lines of evidence, and showing
that some issues under discussion just a few years ago may now be
considered resolved or irrelevant. Other issues remain, however,
and await resolution before the great program of testing the
relativistic Friedmann-Lema\^\i tre model, that commenced in the
1930s, may at last be considered complete. 
\end{abstract}

\section{Introduction}

The search for a well-founded physical cosmology is ambitious,
to say the least, and the heavy reliance on philosophy not
encouraging. When Einstein (1917)
adopted the assumption that the universe is close to homogeneous
and isotropic, it was quite contrary to the available astronomical
evidence; Klein (1966) was right to seek alternatives. But the
observations now strongly support Einstein's homogeneity. Here is a
case where philosophy led us to an aspect of physical reality, even
though we don't know how to interpret the philosophy. Other
aesthetic considerations have been less successful. The steady-state
cosmology and the Einstein-de Sitter relativistic cosmology are
elegant, but inconsistent with the observational evidence we have
now. 

The program of empirical tests of cosmology has been a productive 
research activity for seven decades. The results have greatly
narrowed the options for a viable cosmology, and show that the 
relativistic Friedmann-Lema\^\i tre model passes impressively
demanding checks. My survey of the state of the tests four years
ago, in a Klein 
lecture, was organized around Table~1. I discuss here the
provenance of this table, and how it would have to be revised to
fit the present situation.

The table is crowded, in part because I tried to refer to the main  
open issues; there were lots of them. Some are resolved, but  
an updated table would be even more crowded, to reflect the 
considerable advances in developing new lines of evidence. This
greatly enlarges the checks for consistency that are the
key to establishing any element of physical science. Abundant
checks are particularly important here, because astronomical
evidence is limited, by what Nature chooses to show us and by our
natural optimism in interpreting it.

\begin{figure}[t]
\centerline{\psfig{file=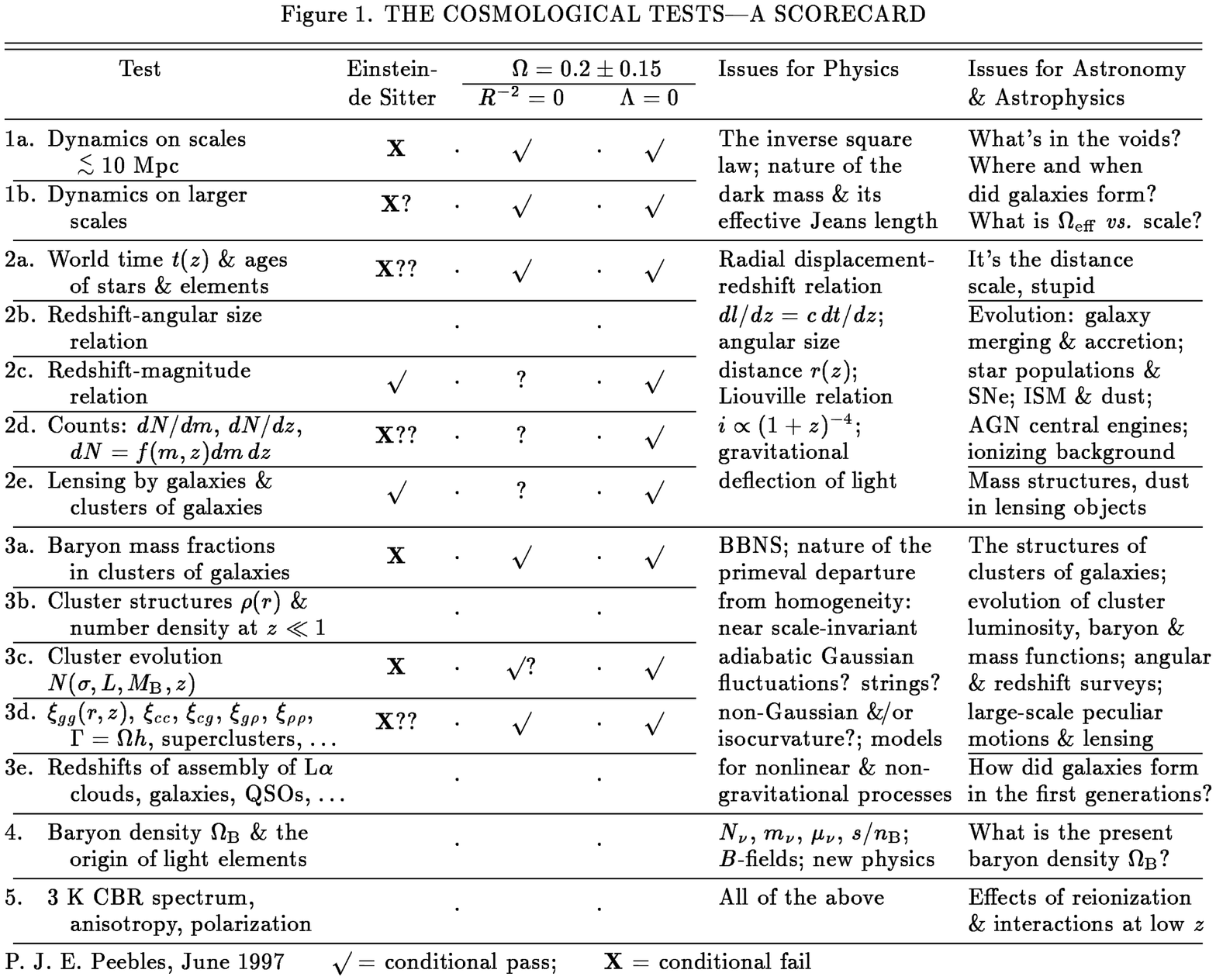,angle=270,width=5.25truein}}
\end{figure}

\section{The Classical Tests}

\subsection{The Physics of the Friedmann-Lema\^\i tre Model}

The second to the last column in the table summarizes elements of
the physics of the relativistic Friedmann-Lema\^\i tre
cosmological model we want to test. 

The starting assumption follows Einstein in taking the observable
universe to be close to homogeneous and isotropic. This 
agrees with the isotropy of the radiation backgrounds and of 
counts of sources observed at wavelengths ranging from radio to
gamma rays. Isotropy allows a universe that is
inhomogeneous but spherically symmetric about a point very close
to us. I think it is not overly optimistic to consider this
picture unlikely, but in any case it is subject to the 
cosmological tests, through its effect on the redshift-magnitude
relation, for example.   

It will be recalled that, if a homogeneous spacetime is described
by a single metric tensor, the line element can be written in the
Robertson-Walker form,
\beq
ds^2 = dt^2 - a(t)^2\left( {dr^2\over 1\pm r^2/R^2}
	+ r^2(d\theta ^2 + \sin ^2\theta d\phi ^2 )\right) .
\label{eq:FLRW}
\eeq
This general expression contains one constant, that measures the
curvature of sections of 
constant world time $t$, and one function of $t$, the expansion 
parameter $a(t)$. Under conventional local physics the de Broglie 
wavelength of a freely moving particle varies as 
$\lambda\propto a(t)$. In effect, the expansion of the universe
stretches the wavelength. The stretching of the wavelengths of
observed freely propagating electromagnetic radiation is measured
by the redshift, $z$, in terms of the ratio of the observed
wavelength of a spectral feature to the wavelength measured at
rest at the source,  
\beq
1 + z = {\lambda _{\rm observed}\over\lambda _{\rm emitted}}
	= {a(t_{\rm observed})\over a(t_{\rm emitted})}.
\label{eq:redshift}
\eeq

At small redshift the difference $\delta t$ of world times at
emission and detection of the light from a galaxy is relatively
small, the physical distance between emitter and observer is 
close to $r=c\delta t$, and the rate of increase of the proper
distance is
\beq
v = cz = H_or, \qquad H_o =\dot a(t_o)/a(t_o),
\label{eq:Ho}
\eeq
evaluated at the present epoch, $t_o$. This is Hubble's law for
the general recession of the nebulae. 

Under conventional local physics Liouville's theorem applies. It
says an object at redshift $z$ with radiation surface brightness 
$i_{\rm e}$, as measured by an observer at rest at the object,
has observed surface brightness (integrated over all wavelengths)   
\beq
i_{\rm o} = i_{\rm e}(1+z)^{-4}.
\label{eq:surf_br}
\eeq
Two powers of the expansion factor can be ascribed to aberration,
one to the effect of the redshift on the energy of each photon,
and one to the effect of time dilation on the rate of reception
of photons. In a static ``tired light'' cosmology one might
expect only the decreasing  photon energy, which would imply 
\beq
i_{\rm o} = i_{\rm e}(1+z)^{-1}.
\label{eq:tired}
\eeq
Measurements of surface brightnesses of galaxies as functions 
of redshift thus can in principle distinguish these
expanding and static models (Hubble \&\ Tolman 1935). 

The same surface brightness relations apply to the 3~K thermal
background radiation 
(the CBR). Under equation~(\ref{eq:surf_br}) (and generalized to
the surface brightness per frequency interval), a thermal
blackbody spectrum remains thermal, the temperature varying as 
$T\propto a(t)^{-1}$, when the universe is optically thin. The
universe now is optically thin at the Hubble length at CBR
wavelengths. Thus we can imagine the CBR was thermalized at
high redshift, when the universe was hot, dense, and optically
thick. Since no one 
has seen how to account for the thermal CBR spectrum under
equation~(\ref{eq:tired}), the CBR is strong evidence for the 
expansion of the universe. But since our imaginations are
limited, the check by the application of the Hubble-Tolman 
test to galaxy surface brightnesses is well motivated (Sandage
1992).   

The constant $R^{-2}$ and function $a(t)$ in
equation~(\ref{eq:FLRW}) are measurable in principle, by the 
redshift-dependence of counts of objects, their angular
sizes, and their ages relative to the present. The metric theory
on which these measurements are based is testable from 
consistency: there are more observable functions than theoretical
ones.   

The more practical goal of the cosmological tests is to
over-constrain  the parameters in the relativistic equation for
$a(t)$,  
\beqa
\left(\dot a\over a\right) ^2 &=&
 {8\over 3}\pi G\rho _t\pm {1\over a^2R^2}\nonumber\\
&=&H_o^2[\Omega _m (1+z)^3 + \Omega _\Lambda +
(1 - \Omega _m -\Omega _\Lambda )(1+z)^2]. \label{eq:H^2}
\eeqa
The total energy density, $\rho _t$, is the time-time
part of the stress-energy tensor. The second line assumes 
$\rho _t$ is a sum of low pressure matter, with energy density
that varies as  $\rho _m\propto a(t)^{-3}\propto (1+z)^3$, and a
nearly constant component that acts like Einstein's cosmological
constant $\Lambda$. These terms, and the curvature term, are 
parametrized by their present contributions to the square of the
expansion rate, where Hubble's constant is defined by
equation~(\ref{eq:Ho}).

\subsection{Applications of the Tests}

When I assembled Table 1 the magnificent program of
application of the redshift-magnitude relation to type Ia
supernovae was just getting underway, with the somewhat mixed
preliminary results entered in line 2c (Perlmutter et al. 1997).
Now the supernovae measurements clearly point to low $\Omega _m$
(Reiss 2000 and references therein), consistent with most other 
results entered in the table.

The density parameter $\Omega _m$ inferred from the application
of equation~(\ref{eq:H^2}), as in the redshift-magnitude
relation, can be compared to what is derived 
from the dynamics of peculiar motions of gas and stars, and from
the observed growth of mass concentrations with decreasing
redshift. The latter is assumed to reflect the theoretical
prediction that the expanding universe is gravitationally
unstable. Lines 1a and 1b show estimates of the density
parameter $\Omega _m$ based on dynamical interpretations of
measurements of peculiar velocities (relative to the uniform
expansion of Hubble's law) on relatively small and large scales.
Some of the latter indicated $\Omega _m\sim 1$. A constraint from
the evolution of clustering is entered in line 3c. The overall
picture was pretty clear then, and now seems well established:
within the Friedmann-Lema\^\i tre model 
the mass that clusters with the galaxies almost certainly is well  
below the Einstein-de~Sitter case $\Omega _m=1$. I
discuss the issue of how much mass might be in the voids between
the concentrations of observed galaxies and gas clouds in \S 4.1.

All these ideas were under discussion, in terms we could
recognize, in the 1930s. The entry in line 2e is based on 
the prediction of gravitational lensing. That was well known in
the 1930s, but the recognition that it provides a cosmological
test is more recent. The entry refers to the multiple imaging of 
quasars by foreground galaxies (Fukugita \&\ Turner 1991). The
straightforward reading of the evidence from this strong lensing
still 
favors small $\Lambda$, but with broad error bars, and it does not
yet seriously constrain $\Omega _m$ (Helbig 2000). Weak lensing
--- the distortion of galaxy images by clustered foreground mass
--- has been detected; the
inferred surface mass densities indicate low $\Omega _m$ (Mellier 
et al. 2001), again consistent with most of the other constraints. 

The other considerations in lines 1 through 4 are tighter, but
the situation is not greatly different from what is indicated in
the table and reviewed in Lasenby, Jones \&\ Wilkinson (2000).  

\section{The Paradigm Shift to Low Mass Density}

Einstein and de~Sitter (1932) argued that the case 
$\Omega _m=1$, $\Omega _\Lambda =0$, is a reasonable 
working model: the low pressure matter term in 
equation~(\ref{eq:H^2}) is the only component one could be sure 
is present, and the $\Lambda$ and space curvature
terms are not logically required in a relativistic expanding
universe. 

Now everyone agrees that this Einstein-de~Sitter model
is the elegant case, because it has no characteristic time to
compare to the epoch at which we have come on the scene. This,
with the perception that the Einstein-de~Sitter model offers the
most natural fit to the inflation scenario for the very early
universe, led to a near consensus in the early 1990s that our
universe almost certainly is Einstein-de~Sitter. 

The arguments make sense, but not the conclusion. For the
reasons discussed in \S 4.1, it seems to me exceedingly difficult
to reconcile $\Omega _m=1$ with the dynamical evidence.

The paradigm has shifted, to a low density cosmologically flat
universe, with  
\beq 
\Omega _m=0.25\pm 0.10,\qquad \Omega _\Lambda = 1 - \Omega _m. 
\label{eq:consensus} 
\eeq 
There still are useful cautionary discussions 
(Rowan-Robinson 2000), but the community generally has settled
on these numbers. The main driver was not the dynamical evidence, 
but rather the observational fit to the adiabatic cold dark
matter (CDM) model for structure formation (Ostriker \&\
Steinhardt 1995). 

I have mixed feelings about this. The low mass density model 
certainly makes sense from the point of view of dynamics
(Bahcall et al. 2000; Peebles, Shaya \&\ Tully 2000; and
references therein). It is a beautiful fit to the new evidence
from weak lensing and the SNeIa redshift-magnitude relation. But
the paradigm shift was driven by a model for structure formation, 
and the set of assumptions in the model must be added to the
list to be checked to complete the cosmological tests. I am
driven by aspects of the CDM model that make me feel uneasy, as
discussed next.   

\section{Issues of Structure Formation}

A decade ago, at least five models for the origin of galaxies and
their clustered spatial distribution were under active discussion
(Peebles \&\ Silk 1990). Five years later the community had 
settled on the adiabatic CDM model. That was in part because
simple versions of the competing models were shown to fail, and
in part because the CDM model was seen to be successful enough to
be worthy of close analysis. But the  
universe is a complicated place: it would hardly be surprising to
learn that several of the processes under discussion in 1990
prove to be significant dynamical actors, maybe along with things
we haven't even thought of yet. That motivated my possibly
overwrought lists of issues in lines 3a to 3e in the table.

I spent a lot of time devising alternatives to the CDM
model. My feeling was that such a simple picture, that we hit on
so early in the search for ideas on how structure formed, could
easily fail, and it would be prudent to have backups. Each of my
alternatives was ruled out by the inexorable advance of the
measurements, mainly of the power spectrum of fluctuations in the
temperature of the CBR. The details can be traced back through
Hu \&\ Peebles (2000). The experience makes me
all the more deeply impressed by the dramatic success of the CDM
model in relating observationally acceptable cosmological
parameters to the measured temperature fluctuation spectrum (Hu
et al 2000 and references therein). 

This interpretation is not
unique. McGaugh (2000) presents a useful though not yet
completely developed alternative, that assumes there is no
nonbaryonic dark matter 
--- $\Omega _{\rm baryon}=\Omega _m\sim 0.04$ --- and assumes a
modification to the gravitational inverse square law on
large scales --- following Milgrom (1983) --- drives structure
formation. But the broad success of the CDM model makes a strong
case that this is a good approximation to what happened as matter
and radiation decoupled at redshift $z\sim 1000$. An update of
Table~1 would considerably enlarge entry~5.  

If the CDM model really is the right interpretation of the CBR
anisotropy it leaves considerable room for adjustment of the
details. I turn now to two issues buried in the table that might
motivate a critical examination of details. 

\subsection{Voids}

The issue of what is in the voids defined by the concentrations
of observed galaxies and gas clouds is discussed at length in
Peebles (2001); here is a summary of the main points. 

The familiar textbook, optically selected, galaxies are strongly
clustered, leaving large regions --- voids --- where the number
density of galaxies is well below the cosmic mean.
Galaxies with 
low gas content and little evidence of ongoing star formation
prefer dense regions; gas-rich galaxies prefer the lower ambient
density near the edges of voids. This is the morphology-density   
correlation. 

The CDM model predicts that the morphology-density correlation
extends to the voids, where the morphological mix swings to favor
dark galaxies. But the observations require this swing to be
close to discontinuous. A substantial astronomical 
literature documents the tendency of galaxies of all known types
--- dwarfs, irregulars, star forming, low surface brightness, and
purely gaseous --- to avoid the same void regions. There are some
galaxies in voids, but they are not all that unusual, apart from
the tendency for greater gas content.  

The natural interpretation of these phenomena is that gravity has
emptied the voids of most galaxies of all types, and with them
drained away most of the low pressure mass. This is is not
allowed in the Einstein-de~Sitter model. If the mass
corresponding to $\Omega _m =1$ were clustered with the galaxies
the gravitational accelerations would be expected to produce
peculiar velocities well in excess of what is observed. That is,
if $\Omega _m=1$ most of the mass would have to be in the voids,
and the morphology-density correlation would have to include the
curious discontinuous swing to a mix dominated by dark galaxies in 
voids. That is why I put so much weight on the dynamical evidence
for low $\Omega _m$. 

At $\Omega _m=0.25\pm 0.10$ (eq.~[\ref{eq:consensus}]) the mass
fraction in voids can be as small as the galaxy fraction. That
would neatly remove the discontinuity. But gravity does not empty
the voids in numerical simulations of the low density CDM model.
In the simulations massive dark mass halos that seem to be
suitable homes for ordinary optically selected galaxies form in
concentrations. This is good. But spreading away from 
these concentrations are dark mass halos that are too small for
ordinary galaxies, but seem to be capable of developing into
dwarfs or irregulars. This is contrary to the observations. 

The consensus in the theoretical community is that the predicted
dark mass clumps in the voids need not be a problem, because
we don't know how galaxies form, how to make the
connection between dark mass halos in a simulation and galaxies
in the real world. The point is valid, but we have some
guidance, from what is observed. Here is an example.  

The Local Group of galaxies contains two large spirals, our
Milky Way and the somewhat more massive Andromeda Nebula. There
many smaller galaxies, most tightly clustered around the two 
spirals. But some half dozen irregular galaxies, similar to the
Magellanic clouds, are on the outskirts of the group.
These irregulars have small velocities relative to the
Local Group. Since they are not near either of the large galaxies 
they are not likely to have been spawned 
by tidal tails or other nonlinear process. 
Since they are at ambient densities close to the cosmic mean
their first substantial star populations would have formed under
conditions not very different from the voids at the same
epoch. In short, these objects seem to prove by their existence
that observable galaxies can form under conditions similar to the
voids in CDM simulations. Why are such galaxies so rare in the
voids? 

\subsection{The Epoch of Galaxy Formation}

Numerical simulations of the CDM model indicate galaxies were
assembled relatively recently, at redshift $z\sim 1$
(eg. Cen \&\ Ostriker 2000). For definiteness in explaining
what bothers me about this I adopt the density parameter in
equation~(\ref{eq:consensus}) and Hubble parameter
$H_o=70$ km~s$^{-1}$~Mpc$^{-1}$. 

The mass in the central luminous parts of a spiral galaxy is
dominated by stars. The outer parts are thought to be dominated
by nonbaryonic dark matter. The circular velocity $v_c$ of a
particle gravitationally bound in a circular orbit in the galaxy
varies only slowly with the radius of the orbit, 
and there is not a pronounced change in $v_c$ at the transition
between the luminous inner part and the dark outer part. The
value of the mean mass density $\rho (<r)$ averaged within a
sphere of radius $r$ centered on the galaxy relative to the
cosmic mean mass density, $\bar\rho$, is
\beq
{\rho (<r)\over \bar\rho } = {2\over\Omega _m}
\left( v_c\over H_or\right) ^2 \sim 3\times 10^5,\quad\hbox{at}\quad
r=15\hbox{ kpc}.
\label{eq:contrast}
\eeq
At this radius the mass of the typical spiral is thought to be
dominated by nonbaryonic dark matter. Why is the dark mass
density so large? Options are that
\begin{enumerate}
\item at formation the dark mass collapsed by a large factor,
\item massive halos formed by the merging of smaller dense
clumps, that formed at high redshift, when the mean
mass density was large, or
\item massive halos themselves were assembled at high redshift.
\end{enumerate}

To avoid confusion let us pause to consider the distinction
between interpretations of large density contrasts in the 
luminous baryonic central regions and in the dark halo of a
galaxy. If the baryons and dark matter were well mixed at high
redshift, the baryon-dominated central parts of the galaxies
would have to have been the result of settling of the baryons
relative to the dark matter. Gneddin, Norman, and Ostriker (2000) 
present a numerical simulation that demonstrates dissipative
settling of the baryons to satisfactory stellar bulges. The
result is attractive --- and hardly surprising since 
gaseous baryons tend to dissipatively settle --- but does not
address the issue at hand: how did the dark matter halos   
that are thought to be made of dissipationless matter get to be
so dense?  

We have one guide from the great clusters of galaxies. The
cluster mass is thought to be dominated by nonbaryonic matter. A 
typical line-of-sight velocity dispersion is 
$\sigma = 750$ km~s$^{-1}$. The mean mass density averaged within 
the Abell radius, $r_{\rm A}=2$~Mpc, relative to the cosmic mean, is 
\beq
{\rho (<r_A)\over \bar\rho } = {4\over\Omega _m}
\left(\sigma\over H_or_{\rm A}\right)^2 \sim 300.
\eeq
Clusters tend to be clumpy at the Abell radius, apparently still
relaxing to statistical equilibrium after the last major mergers,
but they are thought to be close to dynamic equilibrium,
gravity balanced by streaming motions of the galaxies and mass.
This argues against the first of the above ideas: here are 
dark matter concentrations that have relaxed to dynamical support
at density contrast well below the dark halo of a galaxy
(eq.~[\ref{eq:contrast}]).  

The second idea to consider is that a dense dark matter halo is
assembled at low redshift by the merging of a collection of dense
lower mass halos that formed earlier. This is what happens in
numerical simulations of the CDM model. 
Sometimes cited as an example in Nature is the projected
merging of the two Local Group spirals, the Milky Way and the
Andromeda 
Nebula. They are 750~kpc apart, and moving together at
100 km~s$^{-1}$. If they moved to a direct hit they would merge 
in another Hubble time. But that would require either wonderfully
close to radial motion or dynamical drag sufficient to eliminate
the relative orbital angular momentum. Orbit computations 
indicate masses that would be contained in $\rho\propto r^{-2}$
dark halos truncated at $r\sim 200$~kpc, which seems small for
dissipation of the orbital angular momentum. The computations
suggest the transverse relative velocity is comparable to the
radial component (Peebles, Shaya \&\ Tully 2000). That would say
the next perigalacticon will be at a separation $\sim 300$~kpc,
not favorable for merging. Thus I suspect the Local Group spirals
will remain distinct elements of the galaxy clustering hierarchy
well beyond one present Hubble time. If
the Local Group is gravitationally bound and remains isolated the
two spirals must eventually merge, but not on the
time scale of the late galaxy formation picture.

My doubts are reenforced by the failure to observe precursors of 
galaxies. Galaxy spheroids --- elliptical galaxies and the bulges
in spirals that look like ellipticals --- are dominated by old
stars. Thus it is thought that if present-day spheroids were
assembled at $z\sim 1$ it would have been by the merging of star
clusters. These star clusters might have been observable at
redshift $z > 1$, as a strongly clustered population, but they  
are not.  

That leaves the third idea, early galaxy assembly. I am
not aware of any conflict with what is observed at $z < 1$. The
observations of what happened at higher redshift are rich,
growing, and under debate. 

\section{Concluding Remarks}

If the models for cosmology and structure formation on which
Table~1 is based could be taken as given, the only uncertainties
being the astronomy, the 
constraints on the cosmological parameters would be clear. 
The long list of evidence for $\Omega _m=0.25\pm 0.10$, in the
table and the new results from the SNeIa redshift-magnitude 
relation and weak lensing, abundantly demonstrates that we live
in a low density universe. The CBR demonstrates space
sections are flat. Since $\Omega _m$ is small there has to be a
term in the stress-energy tensor that acts like Einstein's 
cosmological constant. The SNeIa redshift-magnitude result favors
a low density flat universe over low density with open geometry 
($\Lambda =0$), at about three 
standard deviations. That alone is not compelling, considering
the hazards of astronomy, but it is an impressive check of what
the CBR anisotropy says.

But we should remember that all this depends on models we are
supposed to be testing. The 
dynamical estimates of $\Omega _m$ in lines 1a and 1b assume the 
inverse square law of gravity. That is appropriate,
because it follows from the relativistic cosmology we are
testing. We have a check on this aspect of the theory, from
consistency with 
other observations whose theoretical interpretations depend on 
$\Omega _m$ in other ways. The CDM model fitted to 
observed large-scale structure requires a value of
$\Omega _m$ that agrees with dynamics. This elegant result was an 
early driver for the adoption of the low density CDM model.
But we can't use it as evidence
for both the CDM model and the inverse square law; 
we must turn to other measures. We have two beautiful
new results, from weak lensing and the redshift-magnitude
relation, that agree with $\Omega _m\sim 0.25$. The latter
does not exclude 
$\Omega _m=\Omega _{\rm baryons}\sim 0.04$; maybe MOND
accounts for flat $v_c(r)$ but does not affect
equation~(\ref{eq:H^2}) (McGaugh 2000). And if we modified
local Newtonian dynamics we might want to modify the physics of
the gravitational deflection of light. 

There are alternative fits to the CBR anisotropy, with new
physics (McGaugh 2000), or conventional physics and an arguably
desperate model for early structure formation (Peebles, Seager
\&\ Hu 2000). They certainly look a lot less
elegant than conventional general relativity theory with the
CDM model, but we've changed our ideas of elegance before. 

In \S 4 I reviewed two issues in structure formation that I think
challenge the CDM model. They may in fact only
illustrate the difficulty of interpreting observations of complex
systems. It's just possible that they will lead us to some radical
adjustment of the models for structure formation and/or
cosmology. I don't give much weight to this, because
it would mean the model led us to the right $\Omega _m$ for
the wrong reason. Relatively fine adjustments are easier to
imagine, of course. With them we must be prepared for fine
adjustments of the constraints on parameters such as $\Lambda$. 

This is quite a tangled web. Progress in applying the many tests, 
including the mapping the CBR temperature and polarization, will
be followed with close attention. 

We have an impressive case for the Friedmann-Lema\^\i tre
cosmology, from the successful fit to the CBR anisotropy and the
consistency of the evidence for $\Omega _m\sim 0.25$ from a broad
range of physics and astronomy. But the cosmological 
tests certainly are not complete and unambiguous, and since 
they depend on astronomy the program is not likely to be
closed by one critical measurement. Instead, we should expect
a continued heavy accumulation of evidence, whose weight will at
last unambiguously compel acceptance. We are seeing the
accumulation; we all look forward to the outcome.  

\acknowledgements

This work was supported in part by the US National Science
Foundation.

\end{document}